\documentclass[a4paper]{spie}  

\usepackage[]{graphicx}
\usepackage{wrapfig,placeins,afterpage}

\usepackage{supertabular}                                                                                                        
                                                                                                                                 
\usepackage[dvips]{color}                                                                                                        
                                                                                                                                 
\title{Simulating Modulated X-ray calibration Sources for future
       X-ray missions, using GEANT4}

\author{
        C.P.~de~Vries\supit{a},
        V.~Fioretti\supit{b},
        J.W.~den~Herder\supit{a},
        E~Schyns\supit{c} and
        S.D.~Pinto\supit{d}
\skiplinehalf
\supit{a} SRON Netherlands Institute for Space Research, 
          Sorbonnelaan~2, 3584~CA~Utrecht, Netherlands \\
\supit{b} INAF Instituto di Astrofisica Spaziale e Fisica Cosmica,
          via~Piero~Gobetti~101, 40129~Bologna, Italy \\
\supit{c} Photonis France S.A.S.,
          Avenue Roger Roncier, 19100 Brive,
          B.P. 520, 19106 Brive Cedex,
          France \\
\supit{d} Photonis Netherlands B.V.,
          Dwazziewegen 2, 9301 ZR Roden, 
          P.O. Box 60, 9300 AB Roden, 
          Netherlands \\
       }

\begin{document}                                                               
\maketitle 

\begin{abstract}
The XIFU X-ray spectrometer instrument on the future
Athena mission needs X-ray calibration sources to calibrate
the gains of the individual detector pixels.
For this purpose, electronically controlled Modulated X-ray
Sources (MXS) are proposed, similar to the calibrations
sources used on the Hitomi spacecraft and which will also
fly on its successor, XARM.
Here we present a simulation package based on the particle
transport GEANT4 toolkit. Using this package, we compute
the results for different targets and window configurations
for the MXS's. The simulations expose the trade-offs to be
made to select the optimum source configuration for the
Athena/XIFU and XARM/Resolve spectrometer instruments.
\end{abstract}

\section{INTRODUCTION}

In astronomy, for future high resolution X-ray spectroscopic imaging, 
micro-calorimeter instruments are planned on the XARM\cite{XARM2017} and Athena\cite{xifu2013} satellites. 
The first time such an instrument was used for astronomical observations was on the
Hitomi\cite{tsujimoto2017} (ASTRO-H) satellite. Its high spectral resolution observations of the Perseus
cluster\cite{perseus} did show that this type of instrument does open up a new window in the 
analysis of hot and energetic objects in the universe. 
Unfortunately this instrument was lost prematurely, due to early failure of the spacecraft.

Micro-calorimeters require regular careful calibration of their detector gains in order to obtain a
reliable energy scale. For this, internal calibration sources are required, which provide a set of 
X-ray emission lines at known energies. In the past, X-ray instruments usually used radio-active sources
for this purpose. Such sources are very stable, but do have the disadvantage of having only a limited
number of lines and being continuously active.

For the Soft X-ray Spectrometer (SXS) instrument on Hitomi, a new type of
electrically activated source was used: the Modulated X-ray
Source (MXS). This source has the advantage that it can provide X-ray calibration lines at multiple energies
and can be pulsed or switched off, to allow astronomical data to be taken during periods without a background
from the calibration source. These sources consist of a photo-cathode, which generates electrons, which are
accelerated towards an anode target, causing X-ray fluorescence lines to be emitted. For details on the Hitomi
MXS sources, see e.g. references \citenum{devries2017},\citenum{devries2012},\citenum{devries2010}.       

The same type of MXS calibration sources is planned for the spectroscopic instruments on the Hitomi successor, XARM,
and the Athena satellite. 

In order to properly design for optimal performance of these sources, and to explore the possibilities in terms
of energy range (potential target materials) and operating conditions, a software package to simulate these sources
is required. In this paper we present such a package based on the
GEANT4\cite{agostinelli2003geant4,allison2006geant4}
particle transport toolkit.

\section{MXS GEANT4 CONFIGURATION}

\begin{figure} 
\centering
\begin{minipage}[t]{0.43\linewidth}
\centering 
\includegraphics[width=9cm,clip]{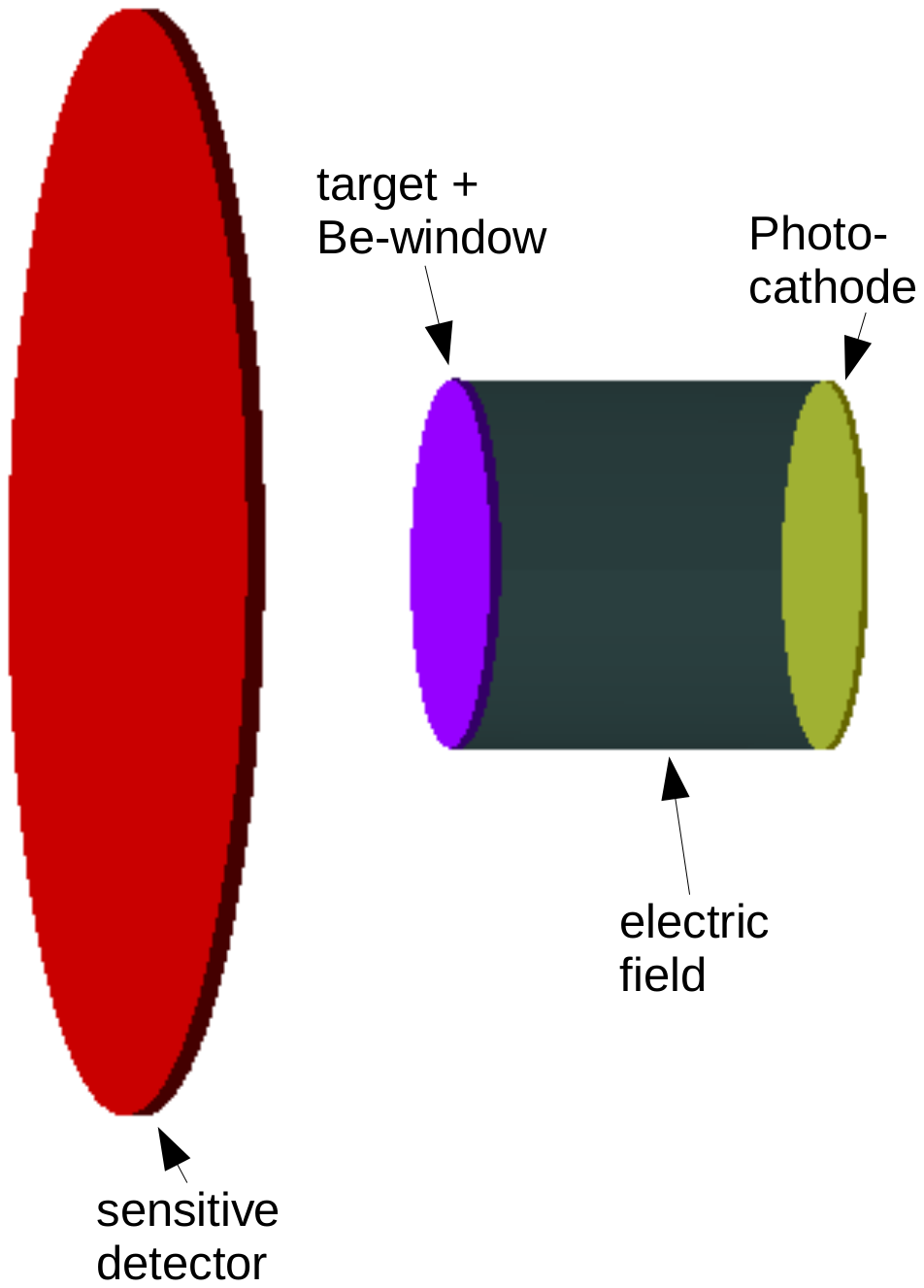}
\caption{\small The direct MXS model. Electrons from the photocathode are accelerated by an electric field towards a 
target, deposited on an, X-ray transparent, Beryllium(Be) window. Fluorescent and Bremsstrahlung X-ray photons exit 
through the Be window and will be detected in the "sensitive detector" volume.}
\label{fig:directMXS}
\end{minipage}
\hspace{1cm}
\begin{minipage}[t]{0.43\linewidth}
\centering 
\includegraphics[width=9cm,clip]{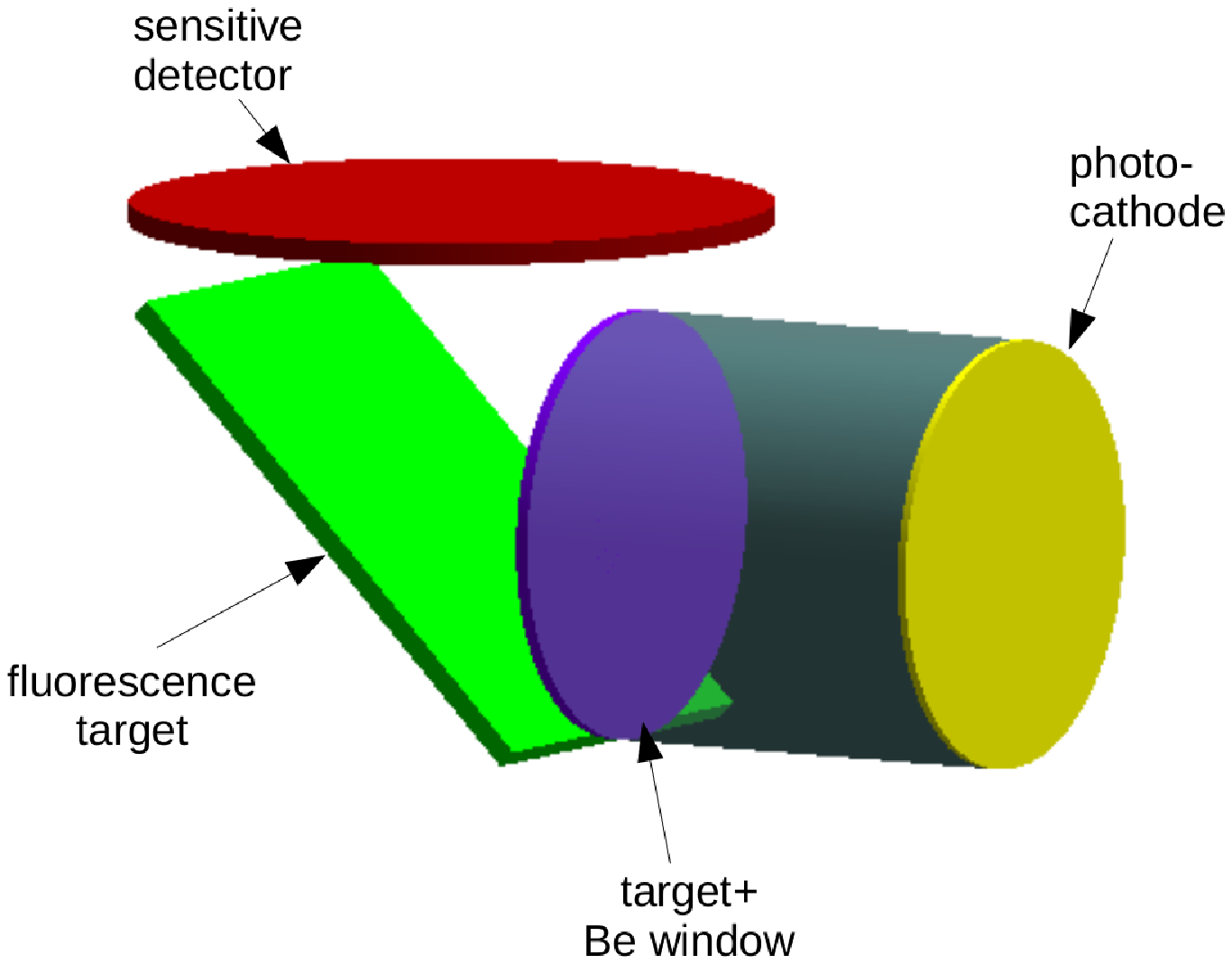}
\caption{\small The indirect MXS model. An external fluorescent target sheet is mounted at a 45 degree angle from the
direct source exit window. The "sensitive detector" is located above the external fluorescent target.}
\label{fig:indirectMXS}
\end{minipage}
\end{figure}

Two types of MXS sources exist: a direct source, in which the X-rays directly coming from the (internal) anode target, 
which is hit by the electrons, are used, and an indirect source, in which the direct X-rays hit an external
fluorescence target, which then generate the X-ray emissions lines. On the MXS an X-ray transparent Beryllium(Be)
window shields the high vacuum inside the source from the outside world. The internal X-ray target is deposited
directly onto the Be window. The direct X-rays leave the MXS through the target and Be window.
Figures \ref{fig:directMXS} and \ref{fig:indirectMXS} show the representation of these
sources in the GEANT4 geometry (the DetectorConstruction class). 

A uniform electric field is defined in between the cathode and the anode (X-ray target). Electrons are generated
uniformly across the cathode and specified to leave the cathode with a negligible 3~eV energy. Subsequently the
electrons are accelerated in the field towards the anode. Any electrons back-scattered from the anode will be
redirected by the electric field, again towards the anode. 

The electron interaction with the anode target and the subsequent propagation of any resulting X-ray photons is
controlled by what is called the GEANT4 "physicslist". Given the MXS low energy range (< 12 keV), in
our MXS "physicslist" only electromagnetic processes for the electron and photon interaction are important. 
For photons, photoelectric and scattering processes are included. For the
electrons, single and multiple scattering processes, ionization (PenelopeIonisationModel) and Bremsstrahlung are
defined. Finally, atomic de-excitation physics adds X-ray fluorescence and Auger processes.

Finally, the "sensitive detector" defines a volume which registers all the particles traversing this volume.
For each registered particle, the nature of the particle is stored: the energy of the particle, 
the location where it originated and the physics process which generated the particle. This way, the performance of the
MXS can be analyzed in detail. 

To speed up processing, the GEANT4 code is started from a python\footnote{https://www.python.org/} script which splits
(using open-mpi \footnote{https://www.open-mpi.org/})
the total number of electrons to be processed over
multiple CPU's (in a linux cluster environment) and at the end combines the output files into a single file which
holds the registered events. 

The MXS model configuration is controlled from an ASCII text file using GEANT4 "G4UIcmd" command definitions. In this
way the MXS configuration can be altered without recompiling the code. Configuration commands include:
\begin{itemize}
\item{magnitude of the accelerating voltage used.}
\item{material and thickness of the different anode targets}
\item{number of consecutive target layers}
\item{thickness of the Beryllium window}
\item{material of the optional external fluorescent target} 
\item{total number of electrons to be processed}
\end{itemize} 

\section{MODEL VALIDATION}

\begin{figure}[h!]
\centering
\begin{minipage}[t]{0.60\linewidth}
\centering
\includegraphics[width=0.80\textwidth,clip]{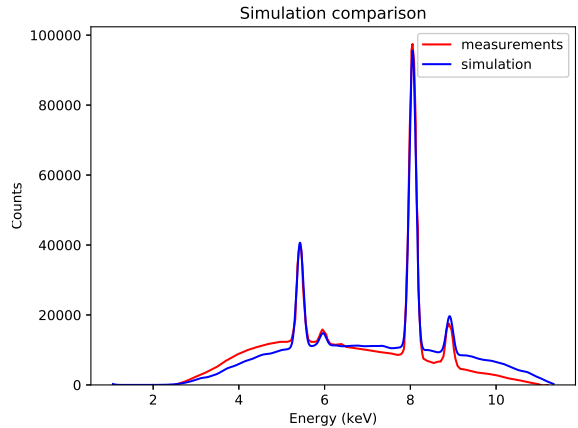}
{\caption{\small Comparison of a GEANT4 simulated MXS spectrum with a spectrum measured from a real Hitomi-SXS MXS.}
\label{fig:comparison}}
\end{minipage}
\end{figure}

Figure~\ref{fig:comparison} shows a comparison between a simulated and a measured spectrum for an MXS identical to the
Hitomi MXS configuration. The simulated spectrum was convolved with the energy resolution of the Silicon Drift Detector(SDD) 
used to measure the spectrum from an actual real source. In addition the simulated spectrum was corrected for absorption
by 8 cm of air, as was the case for the real measurements. The figure shows that the line ratios and the flux in the lines 
with respect to the Bremsstrahlung continuum match quite well. The shape of the Bremsstrahlung is a little different, but
depends quite a bit on the exact match of the accelerating voltage (supposed to be 11.3 kV, but which has some margin of
error (about $\pm 0.3$ kV) for the measurements). In addition, the decrease of SDD efficiency due to the limited
thickness of the silicon of the detector has not been taken into account. Given these uncertainties the simulations match the
measurements quite well.  

The total number of photons generated in the source simulation was 145000 (of which 35000, or 24\% are fluorescent events in 
the Cr-K and Cu-K lines at 5.41 and 8.04 keV respectively). Those events are recorded in the "sensitive detector". 
This volume is located at 0.9~cm from the
Be window and has a radius of 1.5~cm. This means a solid angle of $0.97 \times \pi$~sr. On Hitomi the SXS array had pixels of
0.8~mm square at 87.5~cm from the MXS, corresponding to a solid angle of $2.7 \times 10^{-7} \times \pi$~sr. Thus the count 
rate on a Hitomi SXS pixel is about $2.8 \times 10^{-7}$ part of the simulated count rate on the "sensitive detector".
For 145000 counts this is equivalent to 0.04~counts/pixel.

For the simulations $2 \times 10^{9}$ electrons from the photocathode were used. If this would have been delivered in one
second this corresponds to a current of 0.32~nA. In the measurement setup the X-ray intensity, corrected for absorption
in air, was set at 
$75 \pm 15$ counts/pixel/second. (The relatively large uncertainty stems from the estimated uncertainty in
the measurement geometry about which part of the MXS window was visible from the SDD detector). 
To get 75 counts/pixel/second for the simulations we would need $75.0 / 0.03 \times 0.32$=800~nA. 
In the real experiment 850~nA was measured. 

It can be concluded that the simulation matches the real MXS measurement quite well.

\section{MODEL MXS CONFIGURATIONS FOR XARM AND ATHENA}

\subsection{The standard, higher energy, MXS source}

The GEANT4 model can be used to investigate optimum MXS source configurations and possibilities for the XARM and
Athena missions. For the Hitomi SXS, a type of MXS was used which had an X-ray target of 25~nm of Cr on top of 150~nm
of Cu, deposited on a 300~micron Be window. The design of these target thicknesses was based on very crude estimates
of electron penetration in the targets. Now, with the current proper GEANT4 model, the optimum thicknesses of the targets
can be computed far more accurately. 

\begin{figure}[ht]
\centering
\begin{minipage}[t]{0.70\linewidth}
\centering
\includegraphics[width=0.90\textwidth,clip]{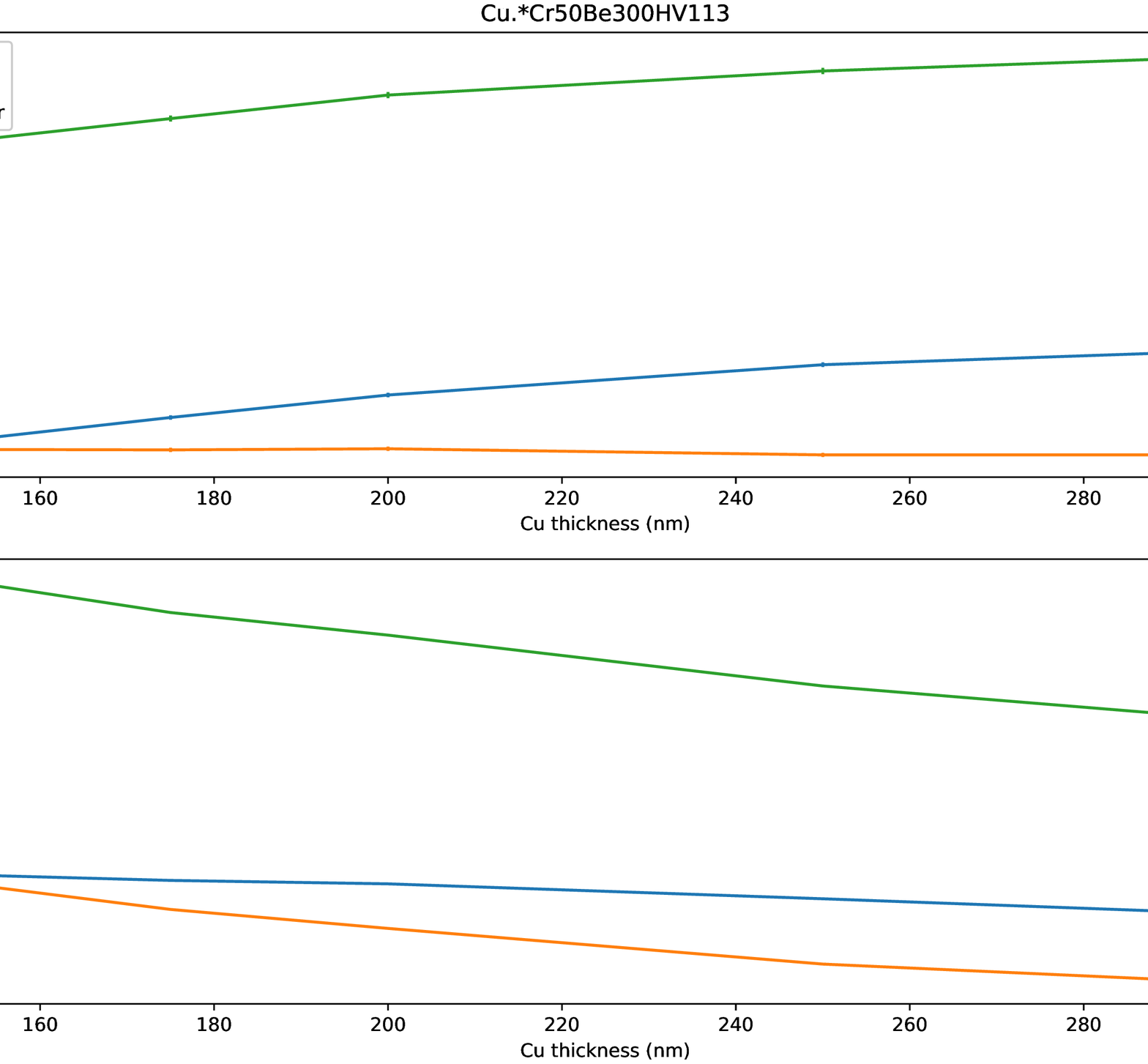}
{\caption{\small Number of X-ray photons in the fluorescent lines when the Cu thickness is increased. All other
parameters (number of generated electrons, High Voltage, Cr thickness etc.) are left the same. The green line shows the total
number of counts in all (Cu+Cr) fluorescent lines.}
\label{fig:Cuchange}}
\end{minipage}
\end{figure}

\begin{figure}[ht]
\centering
\begin{minipage}[t]{0.70\linewidth}
\centering
\includegraphics[width=0.90\textwidth,clip]{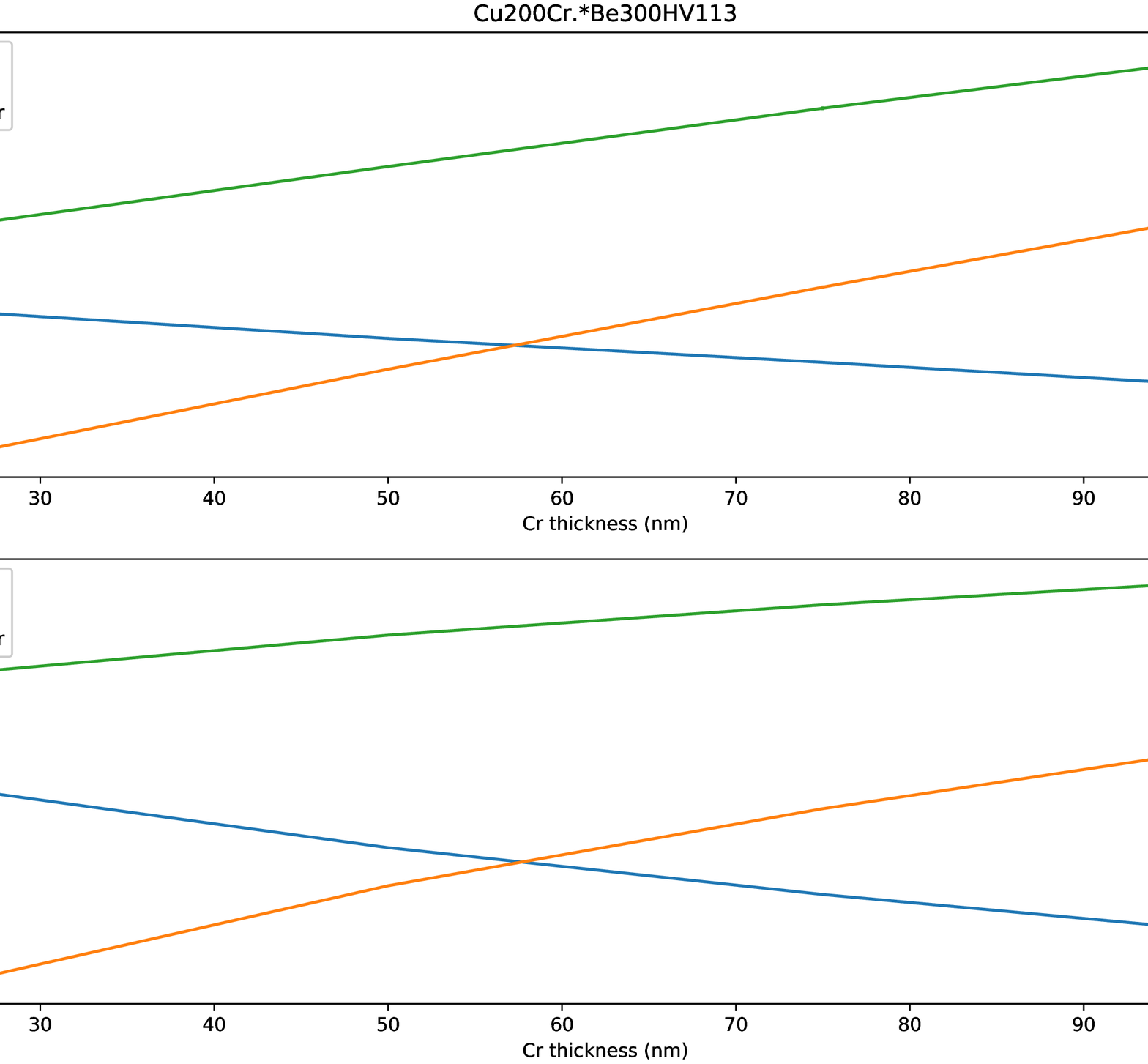}
{\caption{\small Similar as figure~\ref{fig:Cuchange} but now the Cr thickness is increased.}
\label{fig:Crchange}}
\end{minipage}
\end{figure}

Fig.~\ref{fig:Cuchange} shows the number of counts in the Cu and Cr fluorescent lines when the Cu target thickness is
increased. The Cu fluorescent efficiency clearly increases while the Cr efficiency does not suffer.
In Fig.~\ref{fig:Crchange} the Cr thickness is increased. Clearly the Cr fluorescent efficiency increases a lot, but
there is some degradation of the Cu line intensity. This is due to the fact that the Cr is on top of the Cu and high
energy electrons capable of activating the Cu are lost in the Cr layer. 

\begin{figure} 
\centering
\begin{minipage}[t]{0.43\linewidth}
\centering 
\includegraphics[width=8cm,clip]{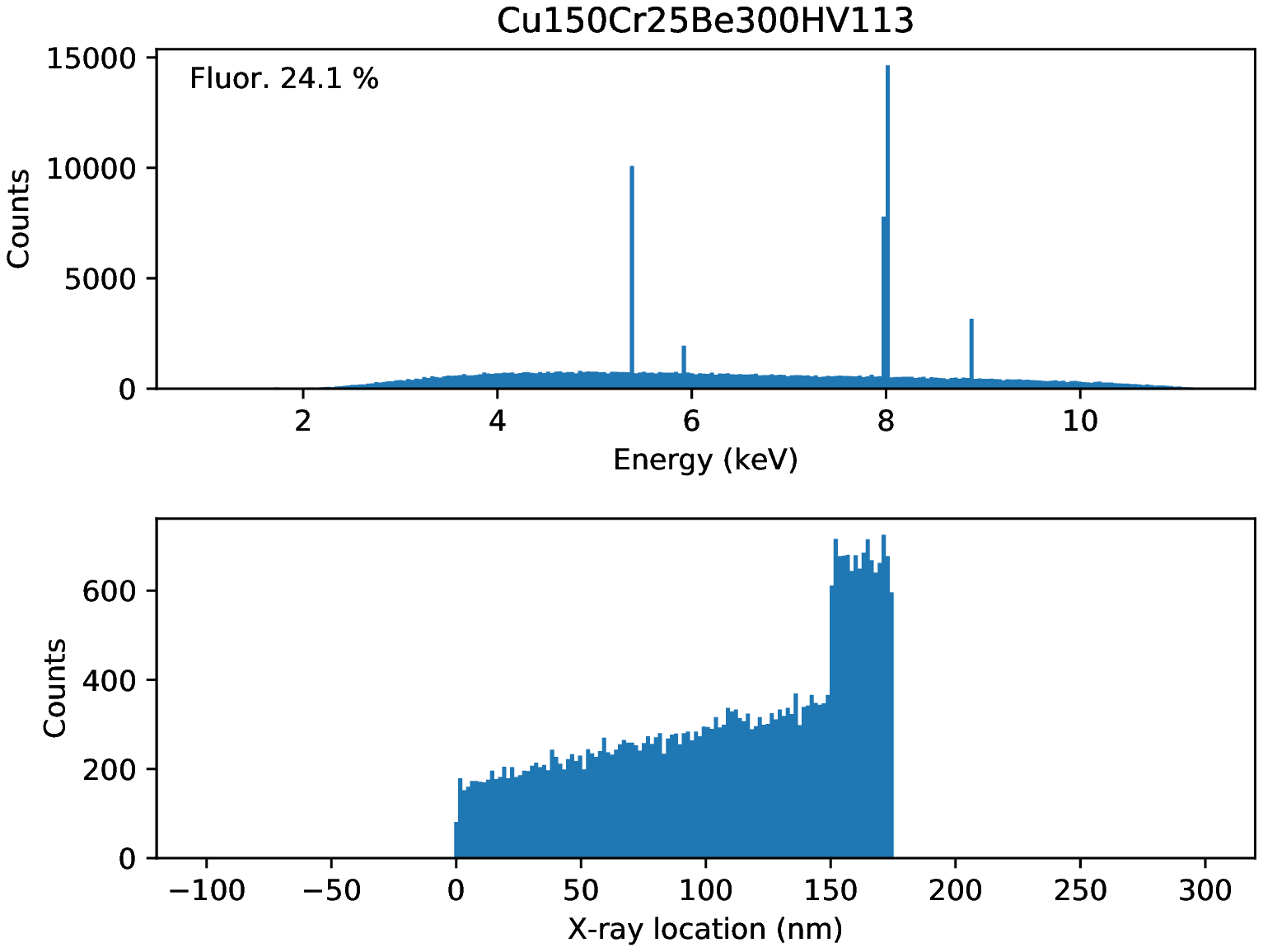}
{\caption{\small Simulated spectrum of the Hitomi MXS source (150~nm Cu and 25~nm Cr).
Bottom plot shows where the fluorescent X-rays lines
originate in the target. The Cr and Cu locations can clearly be identified. This also reflects the penetration of
the electrons in the target.}
\label{fig:HitomiMXS}}
\end{minipage}
\hspace{1cm}
\begin{minipage}[t]{0.43\linewidth}
\centering 
\includegraphics[width=8cm,clip]{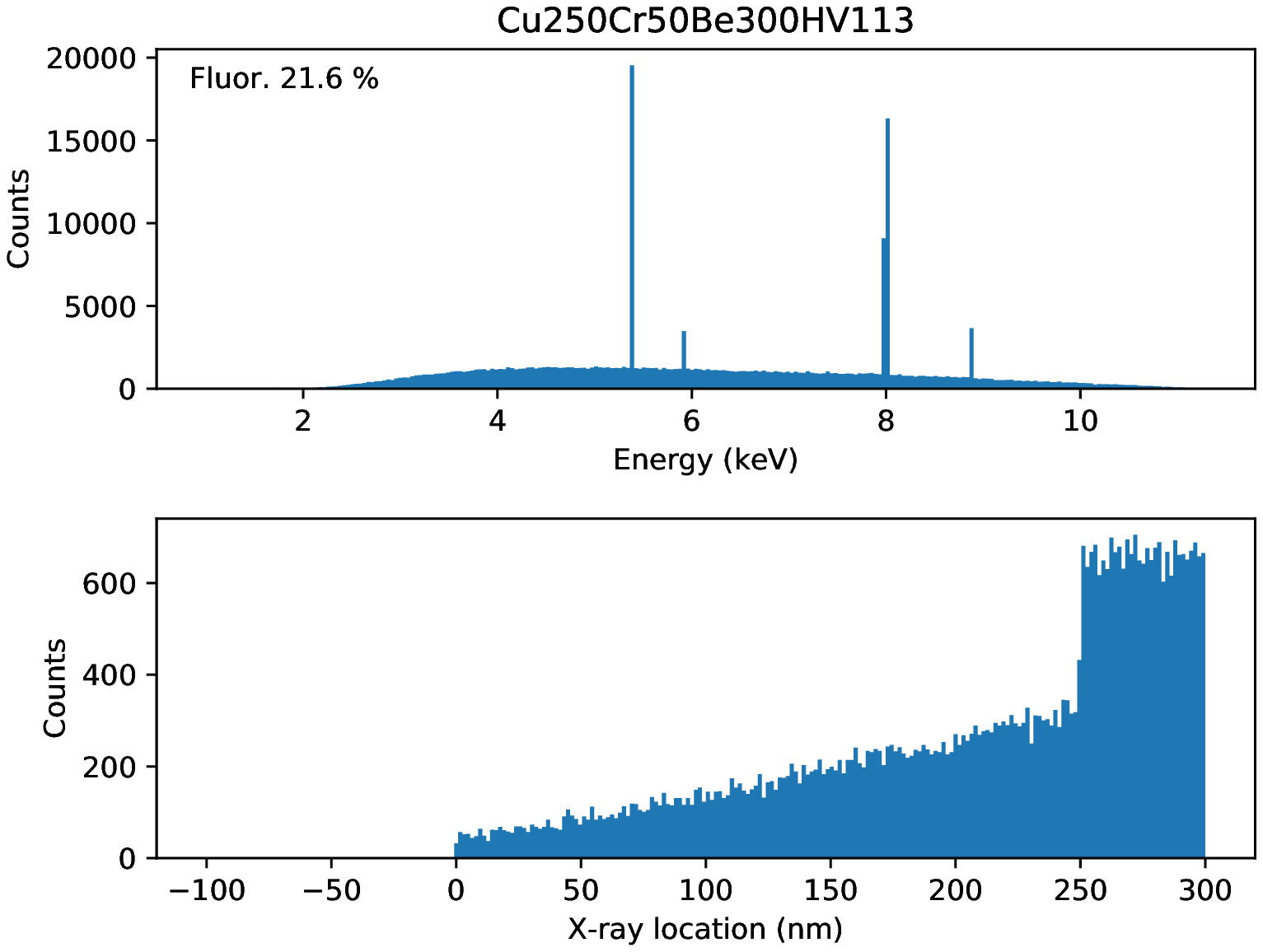}
{\caption{\small Simulated spectrum of the proposed XARM/Athena MXS source with thicker targets (250~nm Cu and 50~nm Cr). 
In the bottom plot it can be seen that
a larger fraction of the electrons is used for fluorescent line generation because in the direction towards the window 
the generated X-rays decline to lower intensities, leading to a total higher source efficiency.}
\label{fig:XARMMXS}}
\end{minipage}
\end{figure}

The optimum performance is for a source with 250~nm Cu and 50~nm Cr targets. Figures~\ref{fig:HitomiMXS} and~\ref{fig:XARMMXS}
compare the spectral performance of the old Hitomi MXS and the proposed optimum (XARM/Athena) MXS. It can be seen, that due
to the thicker targets, the stream of electrons from the cathode will be used more effectively, reflected by the additional decline
of the generated fluorescent X-rays per unit volume in the additional target material towards the Be window.
This leads to a higher source efficiency.  

\begin{figure}[h!]
\centering
\begin{minipage}[t]{0.80\linewidth}
\centering 
\includegraphics[width=8cm,clip]{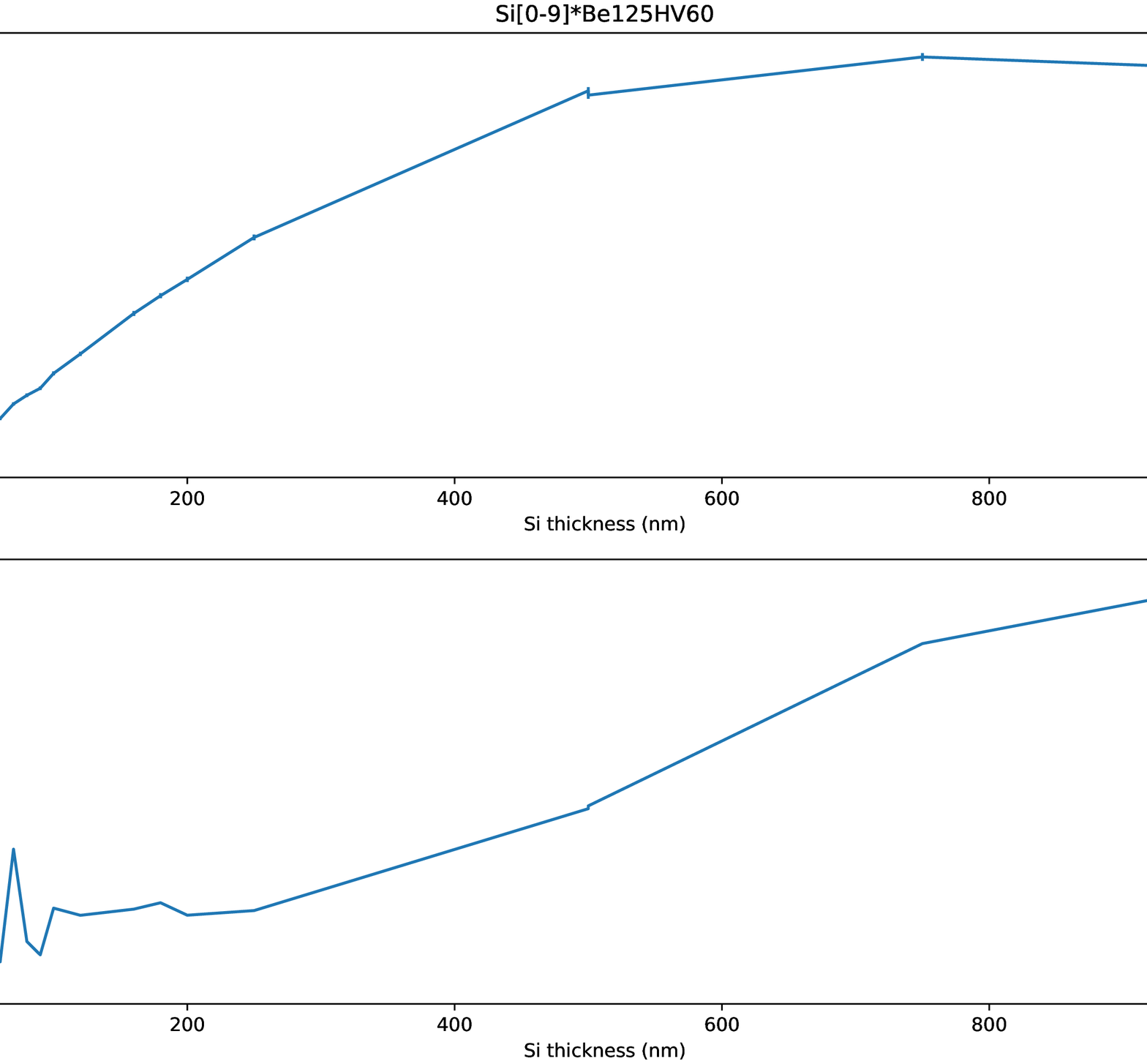}
\caption{\small Simulations of a direct low energy source with Silicon (Si) target. This plot shows the performance as
function of Si thickness. There is an optimum thickness. For thicker targets, internal Si line absorption decreases
the efficiency. The Be window is only 125 micron thin, to prevent excessive absorption of the low energy Si
line by the Beryllium.}
\label{fig:Sithick}
\end{minipage}
\hspace{1cm}
\begin{minipage}[t]{0.80\linewidth}
\centering 
\includegraphics[width=8cm,clip]{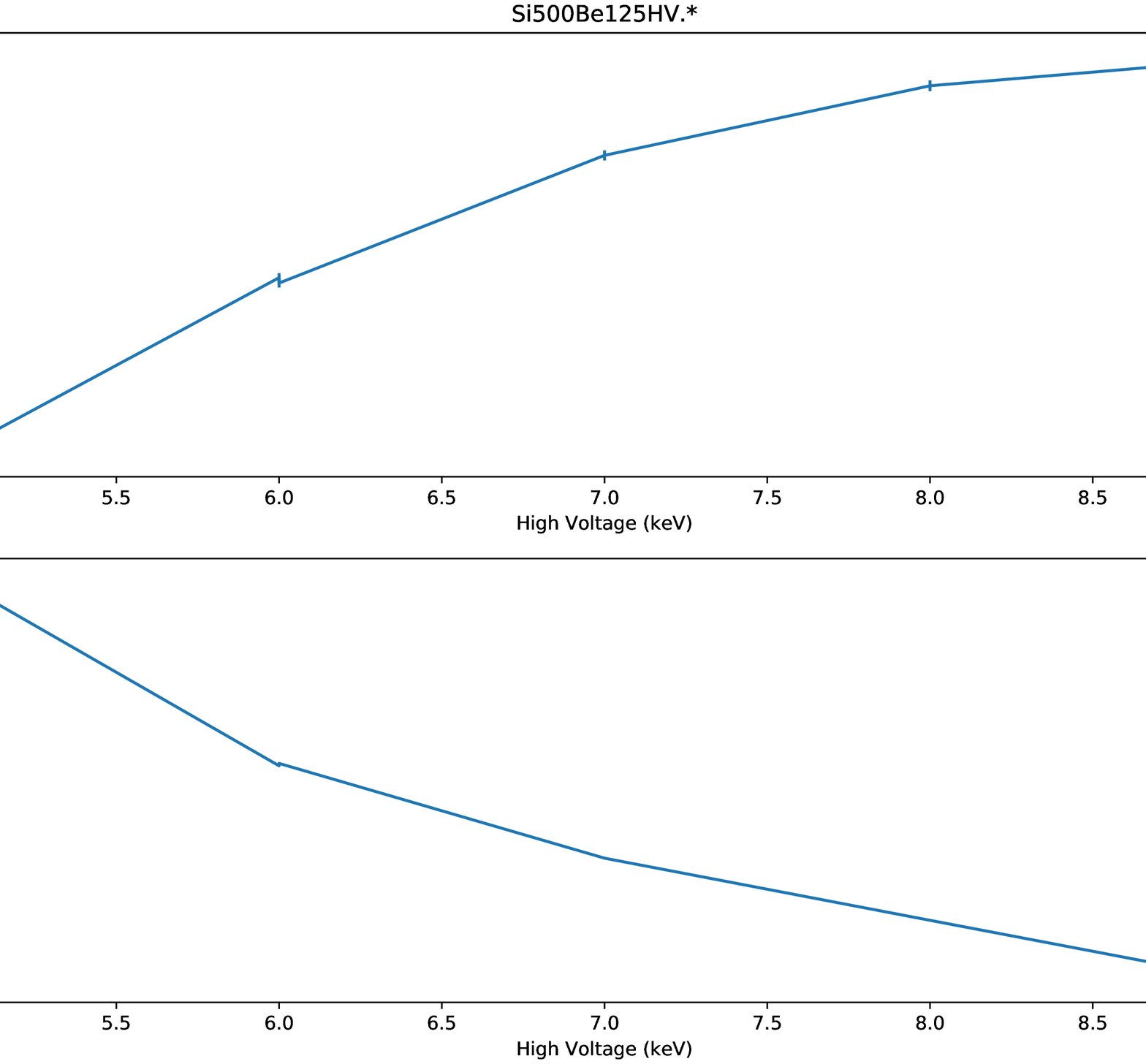}
\caption{\small Performance of a source with a 500~nm Si target as function of the applied accelerating voltage.}
\label{fig:SiHV}
\end{minipage}
\end{figure} 

\begin{figure}[ht]
\centering
\begin{minipage}[t]{0.70\linewidth}
\centering
\includegraphics[width=0.90\textwidth,clip]{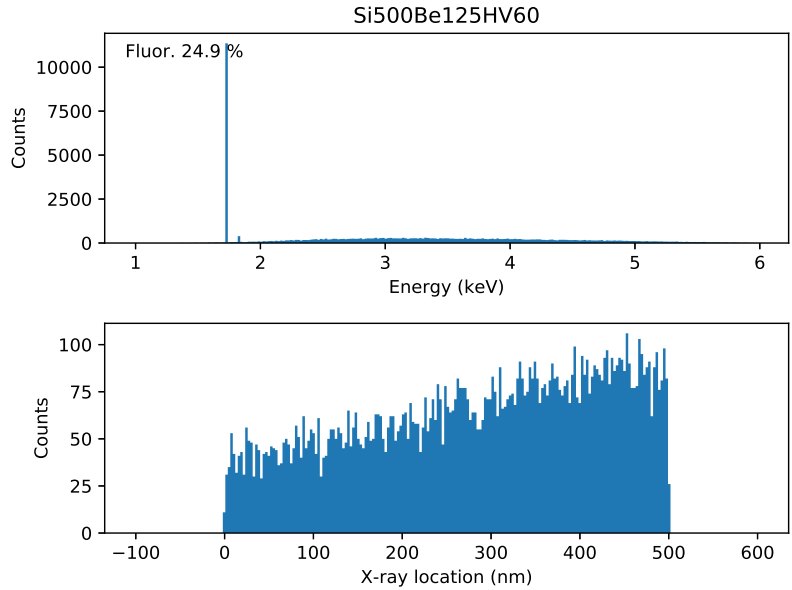}
\caption{\small Spectrum of a direct low energy MXS source with a 500~nm Si target layer and 6.0~kV accelerating voltage.
Not all electrons are 'used' by the target, but increasing the target layer thickness will increase the
internal Si line absorption.}
\label{fig:Sispec}
\end{minipage}
\end{figure}

\subsection{A proposed lower energy MXS source for Athena} 

To check for non-linearities in the detector gain over the entire energy range, calibration emission lines over a broad
range of energies are needed. The Hitomi and proposed XARM direct calibration sources provide lines (Cr,Cu) in the 5~-~9~keV
energy range. For lower energy calibration lines, sources with an external fluorescent target (the indirect sources) 
were used. The problem with these sources is that they have very low efficiency. 
For low energy sources for the Athena XIFU instrument a different approach is investigated.

Problem with low energy targets within the MXS is that the Beryllium window can absorb a significant quantity of the flux. 
The standard 300 micron windows used in the Hitomi/XARM MXS's is clearly too thick for low energy
lines. A compromise is found in a 125 micron Beryllium window and a Silicon (Si) target on top of the window. This combination
allows a source at relatively low energies (1.73 keV) with higher efficiency than an external target. For real low
energies (Al, Mg lines at 1.25, 1.49 keV respectively) an external fluorescent target is still the only option.

Figures~\ref{fig:Sithick}, \ref{fig:SiHV} and \ref{fig:Sispec} show the simulation results. Such a source can be
driven by an accelerating voltage of 6~kV. 

It is planned to test whether the 125 micron Beryllium window can withstand the mechanical stresses and maintain the high vacuum
needed in the MXS's, compared to the proven 300 micron thick windows in the Hitomi/XARM design.

\begin{figure}[h!] 
\centering
\begin{minipage}[t]{0.80\linewidth}
\centering 
\includegraphics[width=0.90\textwidth,clip]{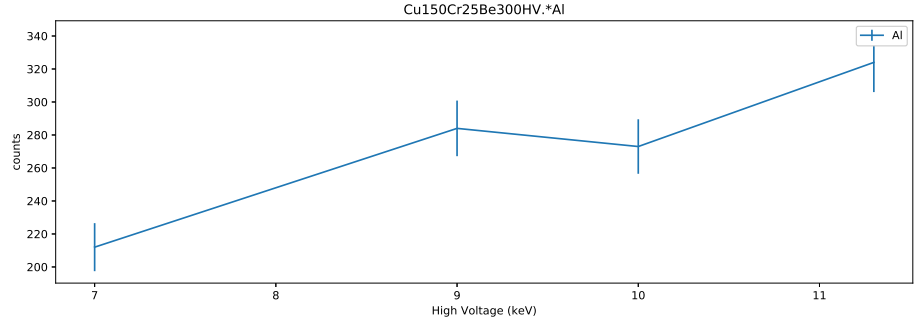}
\caption{\small Standard Hitomi MXS source with an external fluorescent Aluminium (Al) target. Source efficiency as function
of accelerating voltage.}
\label{fig:Fl1HV}
\end{minipage}
\end{figure}

\begin{figure}
\centering
\begin{minipage}[t]{0.80\linewidth}
\centering 
\includegraphics[width=0.90\textwidth,clip]{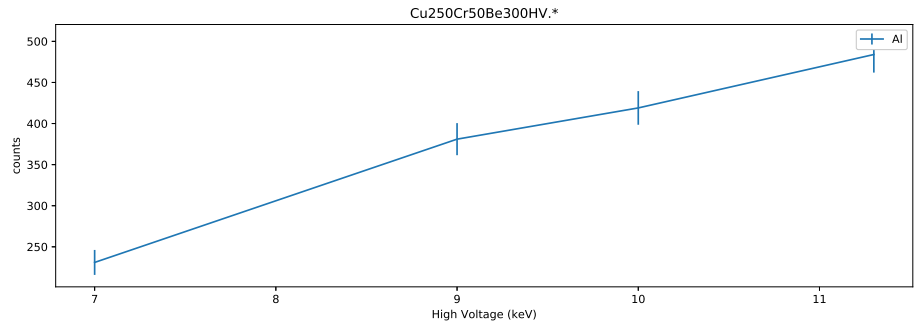}
\caption{\small Proposed XARM MXS source configuration with external Al target. Efficiency as function of accelerating voltage.
Clearly for this MXS source, the efficiency is higher than for the past Hitomi MXS source (Fig.~\ref{fig:Fl1HV}}
\label{fig:FL2HV}
\end{minipage}
\end{figure} 

\subsection{The MXS with external fluorescent target} 

For the lowest energy lines, external fluorescent targets are needed. These fluorescent targets will be exited by
the X-rays (Bremsstrahlung and line emission) from the MXS source. The simulation software can be used to test the
efficiency as function of direct target configuration inside the MXS and accelerating voltage.

Figures~\ref{fig:Fl1HV} and \ref{fig:FL2HV} compare the efficiency of such a source between the Hitomi MXS's
and the proposed XARM configuration. It shows the proposed XARM source is more efficient in generating fluorescent
X-rays and that efficiency keeps increasing with accelerating voltage.

\section{CONCLUSIONS}

The GEANT4 simulation programs are a great tool in designing optimum performance MXS calibration sources. 
The MXS design for the future XARM/Resolve and Athena/XIFU instruments will benefit from this development.
Based on the simulations the XARM/Resolve MXS calibration sources will have improved performance compared to
its Hitomi predecessor, due to a change in target layer thickness. 
In addition the tools allow probing for the best operating voltage, and show the required photo-currents.

\acknowledgements
The research leading to these results has received funding from the European Union's Horizon 2020 Programme
under the AHEAD project (grant agreement 654215).
\endacknowledgements


\bibliography{references}   
\bibliographystyle{spiebib}   

\end{document}